\documentclass[aip,author-year]{revtex4-1}

\usepackage{amsbsy}
\usepackage{amssymb}
\usepackage{amsmath}
\usepackage{graphicx}
\usepackage{color}

\usepackage{multirow}
\def\aap{A \& A}
\def\apj{Ap. J.}
\def\apjl{Ap. J.}

\def\mnras{MNRAS}

\def\prc{Phys. Rev. C}
\def\prd{Phys. Rev. D}
\def\prl{Phys. Rev. Lett.}

\newcommand{\be}{\begin{equation}}
\newcommand{\ee}{\end{equation}}

\newcommand{\tssb}{\tau}
\newcommand{\rmode}{r-mode}
\newcommand{\Iz}{I_{z}}
\newcommand{\fcgw}{f_{\rm CGW}}
\newcommand{\fcgwdot}{\dot{f}_{\rm CGW}}
\newcommand{\fcgwddot}{\ddot{f}_{\rm CGW}}
\newcommand{\scimm}[2]{#1\times10^{#2}}

\newcommand{\frot}{f_{\rm rot}}
\newcommand{\Omegar}{\Omega_r}
\newcommand{\vs}{{\it vs.}}
\newcommand{\eg}{{\it e.g.}}
\newcommand{\ie}{{\it i.e.}}

\newcommand{\Tcoh}{T_{\rm coh}}
\newcommand{\Tobs}{T_{\rm obs}}
\newcommand{\Nseg}{N_{\rm seg}}
\newcommand{\D}{\mathcal{D}}

\begin{document}

\title{Multimessenger observations and the science enabled: Continuous waves and their progenitors, equation of state of dense matter}

\author{D I Jones}

\affiliation{Mathematical Sciences and STAG Research Centre, University of Southampton, Southampton SO17 1BJ, United Kingdom }

\author{K Riles}

\affiliation{Physics, University of Michigan, Ann Arbor, MI, USA}

\begin{abstract}
%\begin{center}

Rotating and oscillating neutron stars can give rise to long-lived \emph{Continuous Gravitational Waves} (CGWs).   Despite many years of searching, the detection of such a CGW signal remains elusive.  In this article we describe the main astrophysical uncertainties regarding such emission, and their relation to the behaviour of matter at extremely high density.  We describe the main challenges in searching for CGWs, and the prospects of detecting them using third-generation gravitational wave detectors.  We end by describing some pressing issues in the field, whose resolution would help turn the detection and exploitation of CGWs into reality.

%\end{center}
\end{abstract}

\maketitle

\tableofcontents

%%%%%%%%%%%%%%%%%%%%%%%%%%%%%%%%%%%%%%%%%%%%%%%%%%%%%%%%%%%%
%%%%%%%%%%%%%%%%%%%%%%%%%%%%%%%%%%%%%%%%%%%%%%%%%%%%%%%%%%%%
%%%%%%%%%%%%%%%%%%%%%%%%%%%%%%%%%%%%%%%%%%%%%%%%%%%%%%%%%%%%
\section{Introduction \label{sect:introduction}}

Next-generation (XG) gravitational wave detectors, such as the proposed facilities of
Einstein Telescope~\citep{bib:EinsteinTelescope} and Cosmic Explorer~\citep{bib:CosmicExplorer}, are expected to
yield detections of compact binary coalescences (CBC) of black holes and neutron stars
in numbers orders of magnitude greater than the cumulative detections to date from
Advanced LIGO~\citep{bib:aligodetector1} and Virgo~\citep{bib:avirgodetector}, as presented in other
articles in this Focus Issue.

Here we consider gravitational sources far longer lived than these
abundant transient phenomena,
%due to cataclysmic mergers in distant galaxies,
namely continuous gravitational waves (CGWs) emitted
by fast-spinning, non-axisymmetric neutron stars in or near to our own galaxy.
This radiation is expected to be nearly monochromatic, albeit with frequency
evolutions and modulations due to the Earth's motion that present signficant
challenges (see section~\ref{sect:methodology}). Unlike the O(100) published
CBC detections to date~\citep{bib:GWTC3}, no CGW detections have been yet made,
and prospects for detection with current advanced detectors are uncertain~\citep{bib:RilesLRR}.

We consider below the prospects for detection with XG detectors, focusing in some detail on two
specific potential emission mechanisms due to 1) neutron star \emph{r-modes}
2) to neutron star ``mountains'', both described in section~\ref{sect:sources}.
We present CGW signal phenomenology in section~\ref{sect:phenomenology}, search methodology in section~\ref{sect:methodology}
and detection prospects in section~\ref{sect:potential}. On the assumption that one or more detections are indeed made,
we discuss in section~\ref{sect:inverse_problem} the \emph{inverse problem} of inferring information concerning
fundamental neutron star physics from what is observable, both gravitationally and electromagnetically.

In some sense, a CGW detection would provide at least the potential for the ultimate multi-messenger source.
Not only would there be an excellent chance for simultaneous detection of gravitational waves and
electromagnetic signals (X-ray, radio and perhaps other wavelength ranges) from the same nearby star,
but one would be able to follow the co-evolution of those signals indefinitely.
In the following, we explore some of those possibilities. This article draws heavily from the
existing literature, including recent detailed review articles on emission mechanism~\citep{bib:LaskyReview,bib:GandG} and
on searches~\citep{bib:SieniawskaBejgerreview,bib:TenorioKeitelSintesreview,bib:PiccinniReview,bib:RilesLRR,bib:WetteReview}

%%%%%%%%%%%%%%%%%%%%%%%%%%%%%%%%%%%%%%%%%%%%%%%%%%%%%%%%%%%%
%%%%%%%%%%%%%%%%%%%%%%%%%%%%%%%%%%%%%%%%%%%%%%%%%%%%%%%%%%%%
%%%%%%%%%%%%%%%%%%%%%%%%%%%%%%%%%%%%%%%%%%%%%%%%%%%%%%%%%%%%
\section{Neutron stars as sources of CGWs \label{sect:sources}}

A NS can emit CGWs in two qualitatively different ways, either through the excitation of one of its \emph{normal modes of oscillation}, specifically the \emph{r-mode instability}, or else through its rotation, if the star supports a so-called \emph{mountain}.  We will review each in turn.

%In the case of normal modes, as long as the oscillation is non-spherical, the mass multipole moments will have a harmomic time dependence, giving GW emission.  In the case of rotation, the star must be deformed away from axisymmetry for there to be GW emission.  

%%%%%%%%%%%%%%%%%%%%%%%%%%%%%%%%%%%%%%%%%%%%%%%%%%%%%%%%%%%%
%%%%%%%%%%%%%%%%%%%%%%%%%%%%%%%%%%%%%%%%%%%%%%%%%%%%%%%%%%%%
\subsection{CGWs from the r-mode instability \label{sect:r-modes}}

NSs are complicated objects, with a rich phenomenology, with a complex set of normal modes to match.  In the idealised case of a perfect fluid, non-rotating star, with no magnetic field and no crust, modelled using Newtonian gravity, the situation is simple, with the balance between fluid pressure and gravity allowing only for the so-call \emph{fundamental} or \emph{f-mode}, and its high frequency overtones, known as the \emph{p-modes}; see \eg\ \citet{cox_80}.  In the realistic case, many more pieces of physics, beyond gravity and ideal fluid pressure, must be included; see, \eg, \citet{andersson_19}.  Roughly speaking, for each such extra piece of physics, there is a family of normal modes, characterised by the relevant restoring force.  For instance, the addition of a magnetic field introduces the \emph{Alfv\'en mode}, the elastic crust introduces \emph{elastic modes}, while compositional stratification of the NS fluid introduces \emph{gravity modes}.

However, all but one of these classes of oscillation are damped rapidly through a combination of viscous processes in the NS fluid, and through gravitational radiation reaction. By ``rapidly'' we mean on timescales \emph{much} shorter than the month-to-year type timescales of relevance to CGWs. For instance, f-modes are damped by gravitational radiation reaction on a timescale of order $0.1$ s; see, \eg, \citet{ak_98}.  The one exception is of great interest, but applies only for rotating stars.  

In the case of rotation, the class of \emph{inertial modes} exists; see, \eg, \citet{lf_99}.  These are oscillations restored by the Coriolis force.  Within this class of oscillations, the \emph{r-modes} are of most interest \citep{aks_99, oetal_98}.  These have the rather unintuitive characteristic that their GW emission tends to make the amplitude of the oscillation grow, a manifestation of the \emph{CFS instability} \citep{fs_78}.  However, the various complicating pieces of physics, such as fluid viscosity, the magnetic field, and the crust, all tend to damp the oscillation, with a strength that depends upon the star's spin frequency and temperature.  The amplifying effect of gravitational radiation reaction increase strongly with frequency \citep{oetal_98}.  

This leads us to the two key outstanding questions in this area. Firstly, how fast must a given NS at a given temperature rotate, so that the amplifying effects of gravitational radiation reaction win out over the other (dissipative) effects?  Secondly, if the mode does indeed go unstable, will it become a source of CGWs?

The answer to the first question requires a careful study of all the dissipative processes, the most important being as follows.  Most obviously, a realistic model of the NS fluid will introduce both \emph{shear viscosity} and \emph{bulk viscosity}.  While shear viscosity will operate throughout the fluid interior, it is like to be most important in a thin \emph{viscous boundary layer} at the NS's crust-core interface \citep{bu_00}.  The strength of this dissipation depends upon the extent to which the elastic (but, crucially, not perfectly rigid) crust itself takes part in the r-mode oscillation \citep{lu_01}.  Clearly, the calculation is a delicate one.  In fact, the situation is even more complex than this: the crust-core interface is not believed to be sharp: the neutron fluid penetrates the inner crust (see, \eg, \citet{cetal_20}).  To complicate matters further, the inner crust itself is believed to form a rather complex \emph{pasta} crystalline structure \citep{netal_22}.  Relevant to all these issues is the nature of the neutron star fluid itself, \ie, its \emph{equation of state} (EoS).  Even the composition is unknown: while it is certain that there is a mixture of neutrons, protons and electrons at lower densities, other species may occur at higher densities, e.g hyperons, pions, or even deconfined quarks; see, \eg, \citet{betal_18_review}.  

\begin{figure}
\begin{center}
\includegraphics[width=12cm]{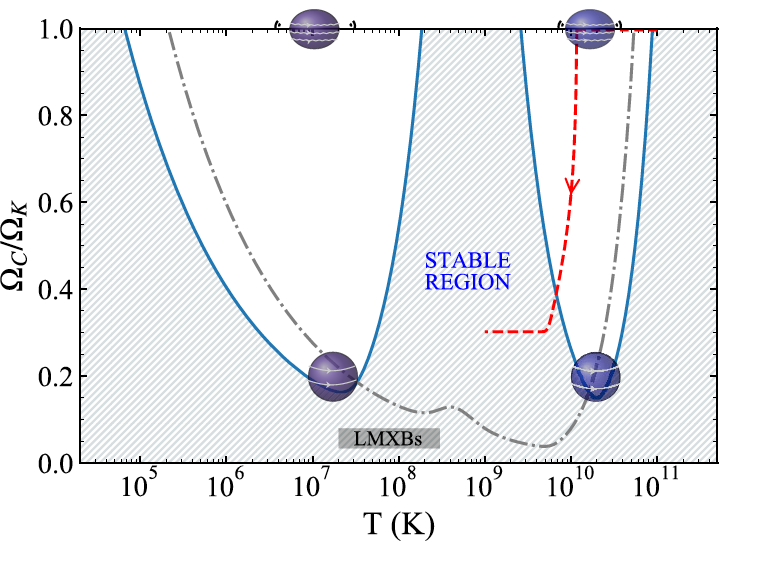}
\caption{The r-mode instability window as computed by (and presented in) \citet{jetal_22}, where special case was taken to model the effects of hyperon bulk viscosity.  The critical spin frequency $\Omega_{\rm C}$ required for instability, normalised to the Keplerian break-up frequency $\Omega_{\rm K}$, is plotted as a function of the stellar temperature $T$.   \label{fig:jetal_22_fig_9}    }
\end{center}
\end{figure}

Work is still ongoing to investigate all these issues, and we will pick out some recent studies.    The dissipation at the shear layer at the crust-core boundary was considered by \citet{netal_14}, who found it was sensitive to input parameters from nuclear physics, including the so-called \emph{symmetry energy} (the energy difference between pure neutron matter and matter with equal numbers of protons and neutrons), and its ``slope" (\ie, its derivative with respect to density).  The role of hyperons on bulk viscosity was examined in \citet{jetal_22} who found that, if present, a hyperonic constituent could stabilise stars over the temperature range $\sim 10^8$ to $\sim 10^9$ K, the relevant range for LMXBs.  This is illustrated in Figure \ref{fig:jetal_22_fig_9}, taken directly from their paper.  Note the existence of both a low and high temperature instability window, a characteristic feature of models that properly incorporate bulk viscosity.

\begin{figure}
\begin{center}
\includegraphics[width=12cm]{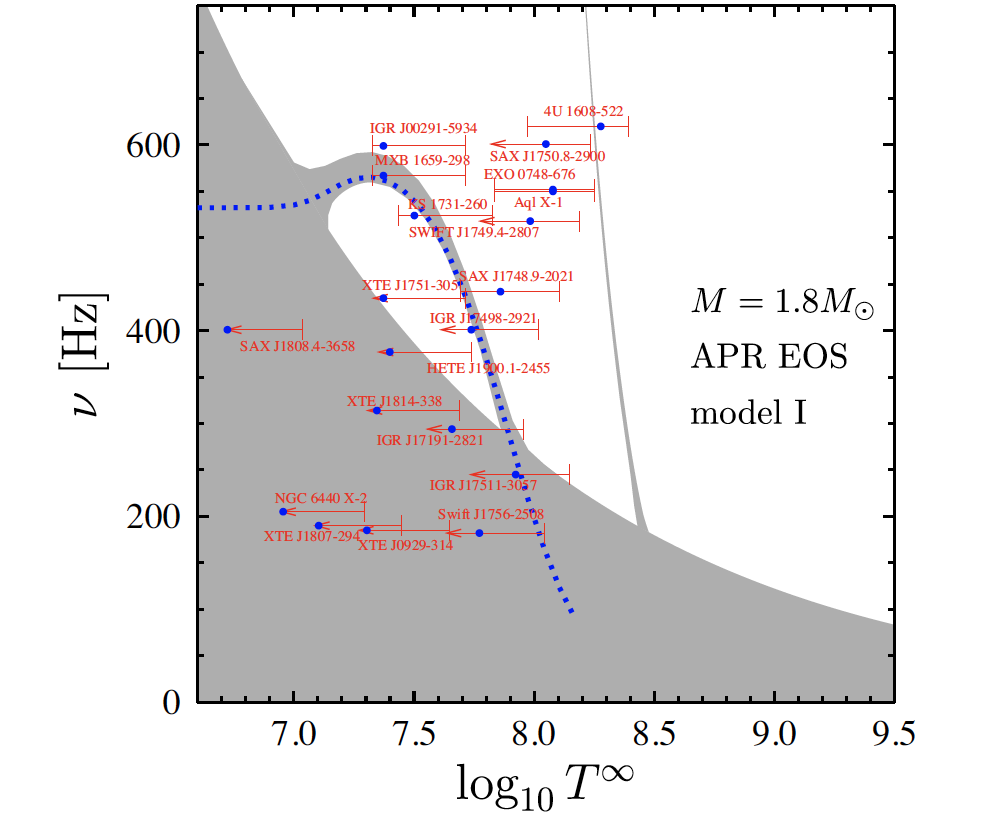}
\caption{The r-mode instability window as computed by (and presented in) \citet{ketal_20}, where the effects of neutron superfluidity were carefully taken into account.  The critical spin frequency $\nu$ varies in a complicated way with temperature $T$ due to an avoided-crossing between the ``normal'' r-mode and its superfluid partner.  \label{fig:ketal_20_fig_4}    }
\end{center}
\end{figure}

The impact of neutron superfluidity on bulk viscosity was considered in \citet{dong_21}, who found while significant, the bulk viscosity was not strong enough to stabilise all known neutron stars.  As well as affecting the transport coefficients, the existence of superfluidity introduces a whole new set of normal modes of oscillation, in which the superfluid and ``normal" (\ie, non-superfluid) components oscillate out of phase.   \citet{ketal_20} have argued that this can lead to so-called ``avoided crossings" between the conventional and superfluid r-modes, leading to resonance features which lead to rather complex and unexpected features in the instability curve.  See Figure \ref{fig:ketal_20_fig_4}.

As is clear from the very different features shown in Figures \ref{fig:jetal_22_fig_9} and \ref{fig:ketal_20_fig_4}, there is not yet agreement on the shape of the instability curve.  Clearly, more work is needed, in which all potentially relevant pieces of physics are simultaneously included, before a consensus can be reached.

The second key issue is, assuming the r-mode does go unstable, what happens next, and what sort of GWs result?  The answer to this depends upon how large the mode grows.  If it grows to large amplitude, a rapidly spinning (perhaps newly born) star might be spun down on a timescale of months to years, making it a CGW source, but a rapidly evolving one.  A schematic trajectory of this sort is shown by the red curve in Figure \ref{fig:jetal_22_fig_9}.   If instead the mode only grows to a small amplitude, it will exert a much weaker spin-down torque on the star.  Indeed, it is perfectly possible that r-modes are active in the accreting LMXB population, explaining their spin rates, or even (but more speculatively) in the young or isolated millisecond pulsars.  R-mode emission from such stars is relatively easily targeted in a CGW search; see Section~\ref{sect:methodology} below.  The study of how large the mode grows necessarily involves non-linear analysis, with energy exchange between modes, but have not been revisited for some time; see, \eg,  \citet{ams_12} and \citet{bw_13}.

%%%%%%%%%%%%%%%%%%%%%%%%%%%%%%%%%%%%%%%%%%%%%%%%%%%%%%%%%%%%
%%%%%%%%%%%%%%%%%%%%%%%%%%%%%%%%%%%%%%%%%%%%%%%%%%%%%%%%%%%%
\subsection{CGWs from mountains \label{sect:mountains}}

% \left(\frac{}{}\right) 

A rotating perfect fluid star will be axisymmetric and emit no GWs.  In order to become a CGW source, such a steadily rotating star needs to be non-axisymmetric, \ie, support a so-called \emph{mountain}.  There are two ways of supporting such a mountain on a NS: using the stars' magnetic field, and using elastic strain.  We will review each.  The size of the mountain can be conveniently parameterised in terms of its \emph{ellipticity}, which is a dimensionless measure of the asymmetry in its moments of inertia, $I_x, I_y, I_z$.  For rotation about $Oz$ we define
\be
\epsilon \equiv \frac{I_1 - I_2}{I_3} .
\label{eq:epsdef}
\ee

In the case of magnetic fields, the fields themselves exert Lorentz forces on the currents that produce them, so all but very special field configurations will generate mountains, with a simple dipolar field giving else to the quadrupolar mass perturbation requires for CGW emission; see \eg, \citet{hsga_08}.  A simple estimate of the ellipticity is given by the ratio of the magnetostatic energy to the star's gravitational binding energy (see, \eg, \citet{jones_02}):
\be
\label{eq:epsilon_B_normal}
\epsilon \sim \frac{B^2 R^3}{GM^2/R} \sim 10^{-12} 
\left(\frac{B}{10^{12} \, \rm G}\right)^2 
\ee
where $M$ and $R$ denote the star's mass and radius, and $B$ is a typical (\eg, at the pole, at the surface) value of the field strength, and we have assumed typical values of $M \sim 1.4 M_\odot$ and $R \sim 10$ km.  This simple estimate is consistent with careful numerical treatments; see, \eg, \citet{lj_09}.  One can obtain larger distortions if suitable field geometries are assumed.  For instance \citet{nanda_23} recently found that the existence of \emph{current sheets}, \ie, infinitely thin surfaces of infinite current density, can lead to larger distortions.

This estimate requires modification in the case where the core protons are superconducting.  The canonical estimate then scales linearly in magnetic field strength \citep{cutler_02}:
\be
\label{eq:epsilon_B_supercon}
\epsilon \sim 10^{-9} \left(\frac{B}{10^{12} \, \rm G}\right)
\ee
but in this case there are not detailed numerical simulations to back this up.  NSs are expected to cool rapidly at birth, so the estimate of equation (\ref{eq:epsilon_B_normal}) applies only for the first hundred or so years of a stars life, after which equation (\ref{eq:epsilon_B_supercon}) should be applied.

\begin{figure}
\begin{center}
\includegraphics[width=12cm]{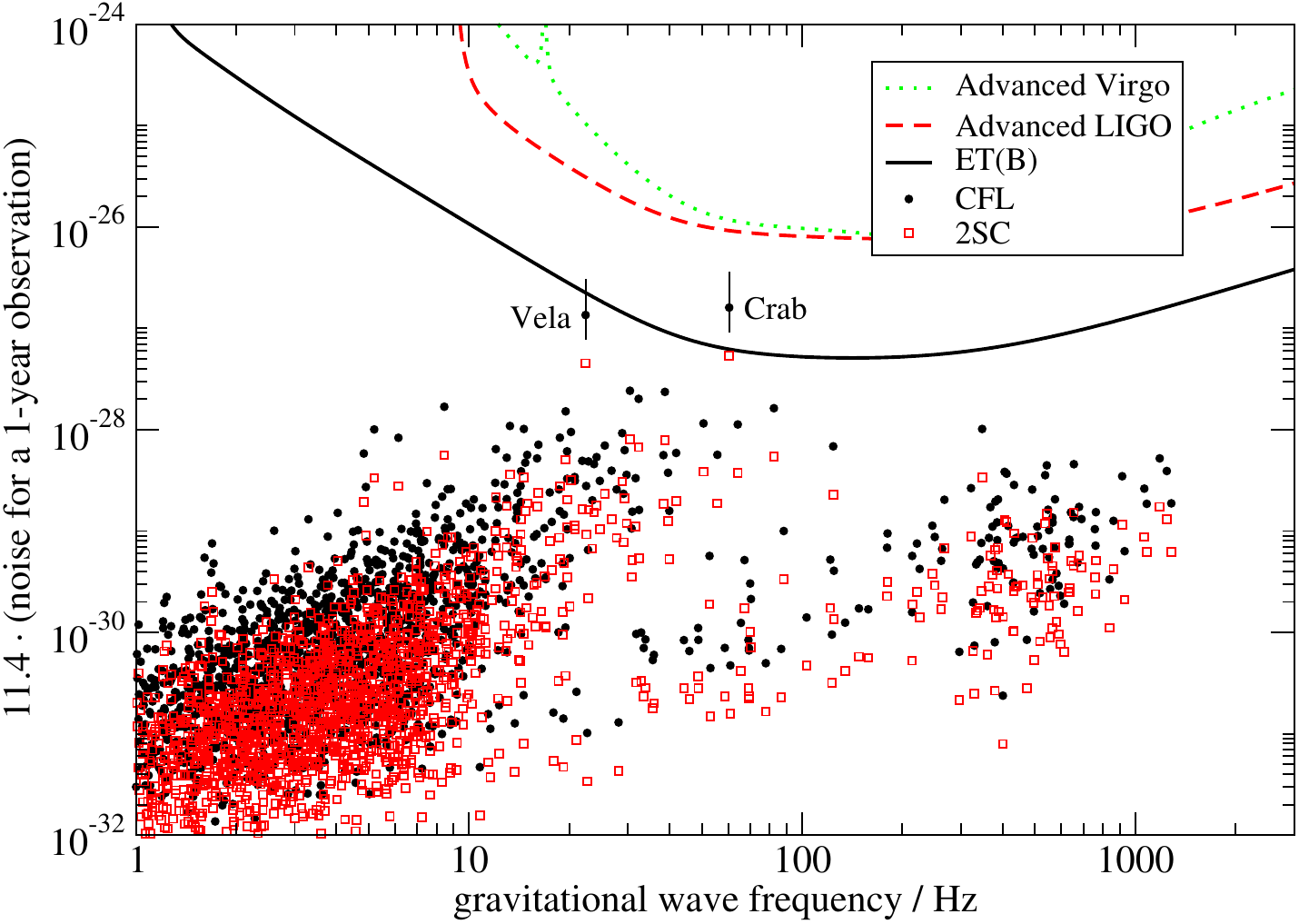}
\caption{Detectability of known pulsars, assuming CFL and 2SC cores, as computed by (and presented in) \citet{gjs_12}.  In the case of CFL cores in the Crab and Vela pulsars, the signals are borderline detectable by the Einstein Telescope.  The vertical error bars spans a plausible range of the quark chemical potential. \label{fig:gjs_12_fig_1}}
\end{center}
\end{figure}

More optimistically from the CGW point of view, larger ellipticities apply if the true EoS of dense matter allows for exotic phases of matter in the neutron star core.  One possible such phase is the \emph{colour-flavour-locked} or CFL phase, where $u$, $d$ and $s$ quarks are involved in pairing \citep{ahs_19}.  It is thought to be inevitable at sufficiently high densities; whether ``sufficiently high" overlaps with actual NS core densities is not known.  Another exotic phase is the so-called \emph{2SC} phase, where only $u$ and $d$ quarks pair \citep{ahs_19}.  As shown in \citet{gjs_12}, if there are sizeable such exotic cores, the \emph{strong nuclear} force (rather than the electromagnetic) is brought into play when producing a magnetic mountain, giving much larger ellipticities than in the conventional purely magnetic case. Indeed, it is possible that some known pulsars may be detectable by third generation GW detectors; see Figure \ref{fig:gjs_12_fig_1}.

In the case of elastic mountains, the situation is more complicated.  One has to distinguish between the \emph{maximum} mountain that a NS can support, given the finite shear modulus $\mu$ and breaking strain $u_{\rm break}$ of its crust, verses the \emph{likely} mountain size, which in turn depends upon what mechanisms drive the mountain formation process.

There has been a lot of work of the issue of maximum mountain sizes.  Simple energy minimisation arguments leads to the following estimate:
\be
\label{eq:epsilon_max_elastic}
\epsilon_{\rm max} \sim \frac{\mu R^2 \Delta R}{GM^2 / R} u_{\rm break} \equiv b u_{\rm break} 
=
10^{-6} \left(\frac{b}{10^{-5}}\right) \left(\frac{u_{\rm break}}{0.1}\right)
,
\ee
where $\Delta R$ is the thickness of the crust; see, \eg, \citet{ja_01}.  The first factor, denoted as $b$ and sometimes known as the \emph{rigidity parameter}, is basically the ratio of the Coulomb binding energy of the crust's crystalline lattice to the total gravitational binding energy of the star.  It is small for typical NSs, $b \sim 10^{-5}$, a reflection of the weakness of the electrostatic crustal force to the gravitational ones \citep{ja_01}.  The breaking strain may be as large as $u_{\rm break} \sim 0.1$, as shown by the molecular dynamics calculations of \citet{hk_09}, although these numerical simulations simulate the crustal material over microscopic timescales; the possibility of plastic creep operating on longer timescales cannot be ruled out, thereby reducing the estimate above; see \citet{ch_10} for a discussion of this issue.

There have been a number of studies that have sought to go beyond the simple estimate of equation  (\ref{eq:epsilon_max_elastic}), solving the full set of equations of elastic equilibrium, generally obtaining results consistent with the estimate.  These include studies in Newtonian gravity \citep{ucb_00, gaj_21}, and in general relativity \citep{jo_13, ga_21}.  There are several technical issues, such as how to treat discontinuities in the shear modulus at the crust-core interface, and how to choose the particular mountain shape \citep{gaj_21, mh_22}, but elastic mountains at the level $10^{-7}$ seem perfectly plausible, in terms of their maximum size.

Again, if one allows for exotic NS cores, the strong nuclear force comes into play, and larger ellipticities are possible.  The presence of  solid quark stars and hybrid quark-baryon stars were examined in \citet{owen_05}, while colour-superconducting cores were considered in \citet{hajs_07}.  These typically give rise to maximum ellipticities about $\sim 1000$ times larger than those of the crust, partly through these exotic phases having a larger shear modulus, and partly through them occupying a larger fraction of the stellar volume.  

The question of \emph{likely} elastic mountain sizes has not received as much attention, in part as it is necessarily a more complex problem.  Put simply, we need to ask what would cause an elastic  mountain to form in the first place; after all, the minimum energy configuration of a spinning star (at fixed angular momentum) will be axisymmetric.

An important step in this direction was taken by \citet{ucb_00}, who considered LMXBs.  They showed that percent-level temperature asymmetries in the accreting NS could lead to ellipticities $\epsilon \sim 10^{-8}$, large enough to explain the sub-Keplerian spin frequencies observed in the Galactic LMXB population.  Physically, this was caused by the slight temperature-dependence of the electron captures that occur as elements of matter are pushed deeper into the crust by the continual accretion.

However, this then poses another question: why should such temperature asymmetries exist in the first place?  An answer to this was provided by \citet{oj_20}, and refined by \citet{hj_23}, who exploited the fact that the magnetic fields threading the NSs will inevitably make their thermal conductivity tensors anisotropic, which in turn guarantees a temperature asymmetry.  It is in fact likely that the LMXB internal magnetic field strength is not large enough for this to produce the required  $\epsilon \sim 10^{-8}$, but nevertheless gives a concrete mechanism for producing an elastic mountain.  See also \citet{setal_20} for a similar analysis of mountain building in LMXBs.

An alternative mountain-building mechanism for accreting stars was suggested in \citet{fhl_18}, who pointed out that as a NS is spun up by accretion, its crust will become elastically stressed, and the resulting ``starquake" may result in a non-axisymmetric fracture.  This suggestion was taken up in more detail by \citet{gc_22}, who showed that this could indeed produce mountains as large as the maximum ones allowed by the crust's finite strength (see equation (\ref{eq:epsilon_max_elastic})).  Their calculation did not take into account the fact that in addition to spinning-up the star, the accretion will tend to replace the primordial NS crust with newly accreted material.  It would be interesting to know if this reduces the rate at which strain is built up, thereby postponing or eliminating the starquake.

There exists a variant of elastic and magnetic mountains that involves the interior neutron superfluid.  This superfluid rotates by forming an array of vortices, and as a NS spins down, the area density of these vortices decreases via a slow radially outward motion, driven by a \emph{Magnus force} acting on the vortices.    These vortices can attach or ``pin'' themselves'' to the crust or, more speculatively, the magnetic flux lines of the proton superconductor.  It is this pinning, and its subsequent release, that provides the standard model of pulsar glitches.  If this pinning is itself non-isotropic, the associated neutron fluid rotation and vortex strains will also be non-axisymmetric.  As first pointed out out in \citet{jones_02}, this would then lead to what might be termed a \emph{Magnus mountain}.  \citet{jones_02} only considered pinning of the vortices on the crustal lattice; the idea was extended to pinning on the magnetic flux tubes in \citet{hap_22}, who found that the GW torque might be strong enough to account for the observed spin-down of the millisecond pulsars.

%%%%%%%%%%%%%%%%%%%%%%%%%%%%%%%%%%%%%%%%%%%%%%%%%%%%%%%%%%%%
%%%%%%%%%%%%%%%%%%%%%%%%%%%%%%%%%%%%%%%%%%%%%%%%%%%%%%%%%%%%
%%%%%%%%%%%%%%%%%%%%%%%%%%%%%%%%%%%%%%%%%%%%%%%%%%%%%%%%%%%%
\section{CGW signal phenomenology \label{sect:phenomenology}}

From the absence of continuous wave detections to date and from realistic assessment of
plausible sources as discussed above, we can expect the first discovered CGW signals to be far weaker than 
the transient signals detected so far. One must integrate the signal over a long duration to observe it
with high statistical significance. Those long integrations in relatively loud noise
require application of assumed signal templates to the data.

Because the Earth spins on its axis once per sidereal day and because it travels once per year
in a nearly circular orbit, CGW signals detected by a ground-based interferometer are
modulated in apparent phase and hence frequency. In addition, there is a sidereal-day amplitude modulation that depends
on the polarization of the signal w.r.t. the interferometer arms.
Searches must account for these resulting modulations. For example,
the Earth's rotation produces a daily relative frequency modulation of $v_{\rm rot}/c\approx10^{-6}$ while the orbital
motion induces a much larger but slower relative frequency modulation of $v_{\rm orb}/c\approx10^{-4}$. 

Explicitly, the time of arrival of a signal at the solar system barycenter, $\tssb(t)$, can be written
in terms of the signal time of arrival $t$ at the detector:
\begin{equation}
\label{eq:phasedefinition}
\tssb(t) \quad \equiv \quad t + \delta t \quad = \quad t - {\vec r_d\cdot\hat k\over c}+ \Delta_{E\odot}+\Delta_{S\odot},
\end{equation}
\noindent where
$\vec r_d$ is the position
of the detector with respect to the SSB, and $\Delta_{E\odot}$ and $\Delta_{S\odot}$ 
are solar system Einstein and Shapiro time delays, respectively~\citep{bib:taylorssb,bib:tempo}.

For illustration, in the following, the mountain model of section~\ref{sect:mountains} will be used
to describe the expected signal model. An \rmode\ model gives a different amplitude dependence on frequency,
but from the perspective of the search, one uses the same form of template.
We treat the signal due to a mountain as that from
an isolated, rotating rigid triaxial ellipsoid, which predicts an intrinsic strain signal amplitude
$h_0$ given by
\begin{eqnarray}
  \label{eq:hexpected}
h_0 & = & {4\,\pi^2G\epsilon\Iz\fcgw^2\over c^4d}  \\
    & = & (\scimm{1.1}{-24})\left({\epsilon\over10^{-6}}\right)\left({\Iz\over I_0}\right)\left({\fcgw\over1\>{\rm kHz}}\right)^2
\left({1\>{\rm kpc}\over d}\right),
\end{eqnarray}
where $\epsilon$ is the equatorial ellipticity defined in equation~\ref{eq:epsdef},
$\fcgw$ is the gravitational wave signal frequency (here, $\fcgw = 2\frot$) and $d$ is the distance to the source.

In equation~\ref{eq:hexpected}, $h_0$ is the amplitude of circularly polarized radiation observable from 
a direction parallel to the stellar spin axis. More generally, for a particular detector, the
potentially observable signal $h$ depends on the orientation of the spin axis with respect to the line
of sight and on the detector's own design and orientation. One parametrization for $h$ as a function of time is
\begin{eqnarray}
\label{eq:cwhdefinition}
h(t) \quad&=&\quad \>F_+(t,\psi)\,h_0{1+\cos^2(\iota)\over2}\,\cos(\Phi(t)) \nonumber \\ 
          & &+ F_\times(t,\psi)\,h_0\,\cos(\iota)\,\sin(\Phi(t)),
\end{eqnarray}
where $\iota$ is the angle between the star's spin direction and the propagation
direction $\hat k$ of the waves (pointing toward the Earth).
$F_+$ and $F_\times$ are the (real) detector antenna pattern response factors
($-1 \le F_+,F_\times \le 1)$ to the $+$ and $\times$ polarizations. $F_+$ and $F_\times$ 
depend on the orientation of the detector and the source, and on 
the polarization angle $\psi$~\citep{bib:cwtargetedS1}. Here, $\Phi(t)$ is
the phase of the signal. A more general signal model with GW emission at both once and twice the rotation frequency is
considered in \citet{bib:JKS}, with effects of free precession addressed in
\citep{bib:JonesAnderssonPrecession1,bib:JonesAnderssonPrecession2,bib:VanDenBroeck,bib:GaoEtal},
and a convenient reparametrization is presented in \citet{bib:JonesParametrization}.

Since emission of gravitational radiation implies energy loss from the stellar source, it is
natural to expect a frequency evolution over time (spin-down in conventional scenarios), such that
the signal phase can be expanded in truncated Taylor form in the
source frame time $\tau$ with respect to a reference time $\tau_0$:
\begin{equation}
\label{eq:phaseevolution}
\Phi(\tau) \quad \approx \quad \Phi_0 + 2\,\pi\left[\fcgw(\tau-\tau_0) + {1\over2}\fcgwdot(\tau-\tau_0)^2 + 
{1\over6}\fcgwddot(\tau-\tau_0)^3\right].
\end{equation}

Detectors are anisotropic in their response functions. In the long-wavelength limit,
Michelson interferometers have an antenna pattern sensitivity with polarization-dependent maxima
normal to their planes and nodes along the bisectors of the arms. As the Earth rotates at angular velocity $\Omegar$ with respect to
a fixed source, the antenna pattern modulation is quite large and polarization dependent via
the functions $F_+(t)$ and $F_\times(t)$ which
depend on the orientation of the detector and the source.

One parametrization of these amplitude response modulations is defined in~\citet{bib:JKS}.
\begin{eqnarray}
  \label{eq:jksone}
  F_+(t)     & = & \sin(\zeta)\left[ a(t)\cos(2\psi)+b(t)\sin(2\psi)\right], \nonumber \\
  F_\times(t) & = & \sin(\zeta)\left[ b(t)\cos(2\psi)-a(t)\sin(2\psi)\right] ,
\end{eqnarray}
\noindent where $\zeta$ is the angle between the arms of the interferometer (nearly or precisely 90 degrees for all major
ground-based interferometers), and where $\psi$ defines the polarization angle of the source wave frame
(\eg, angle between neutron star spin axis projected onto the plane of the sky and
local Cartesian coordinates aligned with its right ascension and declination directions).
The antenna pattern functions $a(t)$ and $b(t)$ depend on the position and orientation of the interferometer on
the Earth's surface, along with the source location and sidereal time; explicit expressions are provided in~\citet{bib:JKS}.

%%%%%%%%%%%%%%%%%%%%%%%%%%%%%%%%%%%%%%%%%%%%%%%%%%%%%%%%%%%%
%%%%%%%%%%%%%%%%%%%%%%%%%%%%%%%%%%%%%%%%%%%%%%%%%%%%%%%%%%%%
%%%%%%%%%%%%%%%%%%%%%%%%%%%%%%%%%%%%%%%%%%%%%%%%%%%%%%%%%%%%
\section{CGW search methodology \label{sect:methodology}}

Searches for CGW signals must take into account the phase/frequency modulations and evolutions embodied
in Eqns.~\ref{eq:phasedefinition}-\ref{eq:phaseevolution} due to detector translational motion, source evolution
and the antenna pattern modulations embodied in equation~\ref{eq:jksone} due to detector orientation modulations.
The templates used to project the weak signal from an integration over months (years) of data must
account for these evolutions. The better the fidelity with which the signal is tracked, the higher the
achievable signal-to-noise-ratio (SNR) for a given signal strength, or for a given SNR threshold, the lower
the achievable signal amplitude. The downside of this improved amplitude sensitivity for the correct template
is the fineness with which one must search in parameter space to find the maximum SNR, to avoid missing
the global maximum.

Depending on what is known about the source, this fineness may present little obstacle, as in the case of a search
for continuous waves from a known pulsar with a well measured ephemeris. On the other hand, for previously unknown sources
in an all-sky search, the necessary fineness for achieving maximum possible SNR makes a search over, say,
a 1-year observation span,
computationally intractable~\citep{bib:RilesLRR}.

These parameter space considerations have led to the following broadbrush categorization of search types
in the CGW literature:
1) {\it Targeted} searches in which the star's position and rotation frequency are known, \ie, known 
radio, X-ray or $\gamma$-ray pulsars;
2) {\it Directed} searches in which the star's position is known, but rotation
frequency is unknown, \eg, a non-pulsating X-ray source at the
center of a supernova remnant; and 3)
{\it All-sky} searches for unknown neutron stars.

The deepest possible searches are those targeted at known pulsars, for which knowledge of the pulsar ephemerides
permit templates that preserve phase fidelity over entire observation runs.
That approach is not feasible in all-sky searches or in directed searches for young sources.
Instead, one must divide long observation periods into time segments ranging, in practice, from minutes to weeks,
depending on the volume of parameter space to be searched. Within each segment, the template search can preserve
fidelity at tractable computational cost. In both directed and all-sky searches, the cost is driven in large part by
the scaling of template count with the segment duration, which typically defines the coherence time of a discrete
Fourier transform (DFT). The Taylor series expansion in equation~\ref{eq:phaseevolution} implies that the fineness
required for searches over $\fcgw$, $\fcgwdot$ and $\fcgwddot$ to maintain an acceptable phase mismatch scale as
$1/\Tcoh$, $1/\Tcoh^2$ and $1/\Tcoh^3$, respectively. As a result, the number of different templates to search
for a given volume of frequency and frequency derivative space explodes for large values of $\Tcoh$~\citep{bib:RilesLRR}.

An additional, similar consideration for all-sky searches is that the fineness with which sky locations must be searched
also depends on coherence time and on the frequency itself because of the sky-dependent Doppler frequency shift.
A rule of thumb~\citep{bib:cwallskyS2} for necessary angular resolution, driven mainly by the Earth's orbital motion, is the following:
\begin{equation}
\delta\theta \quad \approx \quad \scimm{9}{-3}\>{\rm rad}\>\left({30\>{\rm minutes}\over T_{\rm coh}}\right)
\left({300\>{\rm Hz}\over \fcgw}\right),
\end{equation}
where resolutions in both right ascension and declination are affected by this scaling.

%For all-sky searches, it is easy to exhaust all available computational resources by choosing a too-long coherence
%time. Hence one must compromise via segmentation, leading to reduced strain amplitude sensitivity.
The most common choice of search type when using  segmentation is what is known
as a {\it semi-coherent} search, as opposed to the fully coherent
search feasible for known pulsars. In a semi-coherent search, phase fidelity is preserved within each segment, but
is allowed to vary freely from one time segment to the next~\citep{bib:StackSlide2}. An intermediate variation receiving
more use in recent years is known as {\it loosely coherent}~\citep{bib:loosecoherence}, in which the detection statistic
favors signal templates with approximate phase consistency across neighboring segments. While fully coherent searches,
when computationally feasible, have strain sensitivities that scale with observation time $\Tobs$ as $1/\Tobs^{1/2}$, 
semi-coherent searches yield sensitivities that typically scale as $1/\Tobs^{1/4}$. An alternative but equivalent perspective is that for
a fixed segment length ($\Tcoh$), the amplitude sensitivity typically scales as $1/\Nseg^{1/4}$, where $\Nseg$ is the number of segments.
The scaling of loose coherence sensitivity with segmentation lies
between these and depends on a phase resolution parameter that can be tuned
to trade off sensitivity for computational tractability.  See \citet{wette_12} and \citet{ps_12} for a discussion of these issues.

Because of the large variation in methodology, including choices in $\Tcoh$ and other search parameters, it is helpful
to characterize search methods in terms of a ``bottom line'' for comparisons. A common figure of merit is
the {\it sensitivity depth}~\citep{bib:SensitivityDepth}, defined 
with respect to a given intrinsic detector strain
amplitude spectral noise density (square root of power spectral noise density $S_h$):
\begin{equation}
  \D \equiv {\sqrt{S_h}\over h_0},
\end{equation}
\noindent where $h_0$ is the quantity of interest, typically the 90\%\ or 95\%\ upper limit on a strain
amplitude. Numerical values (units: Hz$^{-1/2}$) range for templated searches from $\sim$1000 for
targeted searches of $\sim$2 years down to $\sim$20 for the most sensitive all-sky search for CGW signals
in unknown binary systems~\citep{bib:RilesLRR}.

A different measure of sensitivity, specific to known pulsars, makes the optimistic (and undoubtedly wrong, in many cases) assumption that the inferred rotational spin-down from measured ephemerides
and implied rotational kinetic energy loss can be entirely attributed 
to gravitational wave energy emission. Under this ``gravitar'' assumption, one can derive
the following spin-down limit $h_{\rm spin-down}$ from equation~\ref{eq:hexpected}:
\begin{eqnarray}
\label{eq:spin-downlimit}
h_{\rm spin-down} & = & {1\over d}\sqrt{-{5\over2}{G\over c^3}\Iz{\fcgwdot\over\fcgw}} \nonumber \\
& = &  (\scimm{2.6}{-25})
\left[ {1\>{\rm kpc}\over d} \right]\!
  \left[
    \left({1\>{\rm kHz}\over\fcgw}\right)\!
    \left({-\fcgwdot \over10^{-10}\>{\rm Hz/s}}\right)\!
    \left({\Iz\over I_0}\right)\!
    \right]^{1\over2}\!\!,\quad\>\>\>
\end{eqnarray}
\noindent where $\fcgwdot$ is the signal frequency first derivative.

Thus for any pulsar with a measured frequency, spin-down and
distance $d$, one can predict the maximum possible detectable strain amplitude
on Earth allowed by energy conservation. Then comparing this nominal maximum
possible amplitude with what is detectable at the star's nominal signal frequency
for a set of detectors gives an optimistic measure of possible detection prospects.
As noted above, however, any rotational energy loss due to magnetic dipole radiation emission,
to pulsar winds or to other dissipative mechanisms, can dramatically outweigh the actual GW energy loss.

The higher the necessary equatorial ellipticity $\epsilon$ to account for all of the observed energy loss,
the more unlikely it is that the loss in indeed primarily gravitational.
For known pulsars, it is hence useful to translate the spin-down limit into a corresponding
upper limit on ellipticity $\epsilon_{\rm spin-down}$, again, under the gravitar assumption:
\begin{eqnarray} 
  \epsilon_{\rm spin-down} & = & \left[{5\over32\,\pi^4}{c^5\over G\Iz}{-\fcgwdot\over\fcgw^5}\right]^{1/2} \nonumber \\
                         & = & 2.4\times10^{-7}
                               \left[\left({1\> {\rm kHz}\over\fcgw}\right)^5 \left({-\fcgwdot\over10^{-10}\>{\rm Hz/s}}\right) \left({I_0\over I_{zz}}
                               \right) \right]^{1/2}
  \label{eq:maxeps}
\end{eqnarray}
The $\fcgw^{-5/2}$ scaling in this formula means that maximum ellipticities for older millisecond
pulsars at high frequencies consistent with observation
tend to be orders of magnitude lower than those of young pulsars at low
frequencies, as illustrated in the next section. Unfortunately, millisecond pulsars are all old and
likely highly annealed, leading to expected ellipticities that are still lower.

%%%%%%%%%%%%%%%%%%%%%%%%%%%%%%%%%%%%%%%%%%%%%%%%%%%%%%%%%%%%
%%%%%%%%%%%%%%%%%%%%%%%%%%%%%%%%%%%%%%%%%%%%%%%%%%%%%%%%%%%%
%%%%%%%%%%%%%%%%%%%%%%%%%%%%%%%%%%%%%%%%%%%%%%%%%%%%%%%%%%%%
\section{Expected XG discovery potential \label{sect:potential}}

The CGW discovery potential of XG detectors will depend, of course, on their noise levels $\sqrt{S_h}$, but also
on the search types, observation spans and particular algorithms. In the following, we consider two representative
search types and what we hope to be realistic sensitivity parametrizations.

First, consider targeted searches, for which we can use an exact ephemeris. A 95\%\ confidence level upper
limit for a Bayesian targeted search (pulsar spin axis orientation unknown) with a single detector is shown
in Gaussian noise simulations to be approximately~\citep{bib:DupuisWoan}:
\begin{equation}
  h_{95\%} \sim 11\sqrt{S_h/\Tobs},
  \label{eq:targetedsens}
\end{equation}
\noindent where for $\Tobs$ $\sim$ 1 year, the corresponding single-detector sensitivity depth is $\sim$500 Hz$^{-1/2}$.
Since for semi-coherent all-sky searches the strain sensitivity scales as $\Tobs^{-1/4}$, 
longer observation periods are less helpful than in targeted searches for improving sensitivity.

In the following comparisons, the Einstein Telescope design configuration known as
ETC (3 detectors in a triangular configuration) are used for
assessing detection prospects, where for targeted searches a 5-year run is assumed (neglecting deadtime), using
equation~\ref{eq:targetedsens}, and where for all-sky searches a sensitivity depth of
50 Hz$^{-1/2}$ is used~\citep{bib:RilesLRR}.

\begin{figure}[t!]
\begin{center}
\includegraphics[width=13.cm]{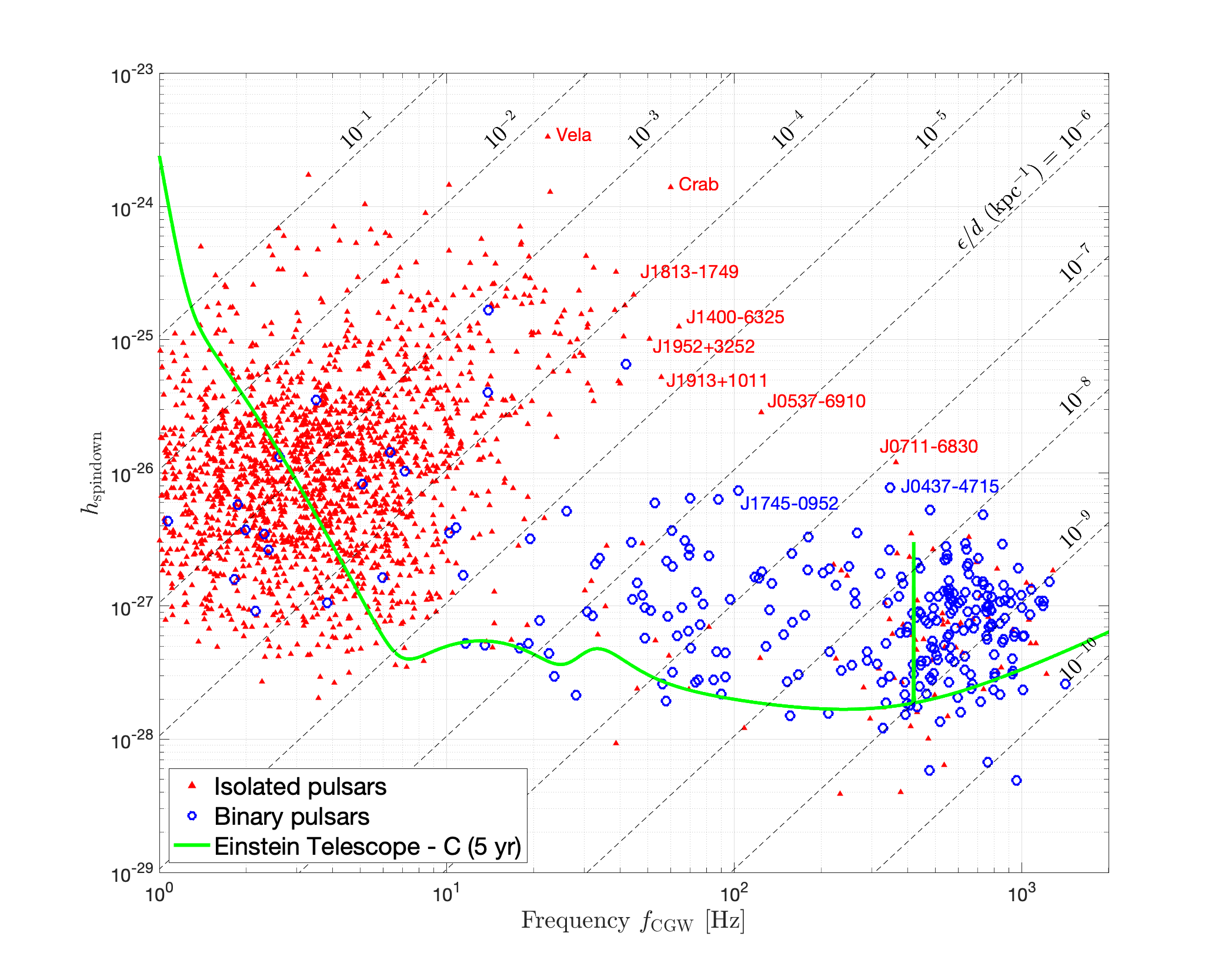}
\caption{Nominal expected GW frequencies and nominal strain spin-down limits for known
pulsars. Closed triangles indicate isolated stars. Open circles
indicate binary stars. The green curve indicates the expected sensitivity for
a 5-year, 3-detector Einstein Telescope data run in the ETC configuration. Dashed diagonal lines correspond to particular
quotients of ellipticity over distance.
A subset of pulsars of particular interest are labeled on the figure.}
\label{fig:fvshspindown}
\end{center}
\end{figure}

Figure~\ref{fig:fvshspindown} shows the spin-down gravitational wave strain limit (gravitar assumption)
\vs\ nominal GW signal frequency (2$\frot$)
for known pulsars\footnote{From ATNF Catalog (release V1.66)~\citep{bib:ATNFdb}}
with well measured ephemerides. Also shown are corresponding contours of constant implied values of $\epsilon/d$, under the gravitar assumption, for reference. The green curve shows the nominal targeted-search
sensitivity for the ETC configuration. One sees that most millisecond pulsars are, in principle, detectable
under the gravitar assumption.
%and that the ellipticities needed for detection are at least in a maximum possible value
%range for stars as close as $\sim$1 kpc.

\begin{figure}[t!]
\begin{center}
\includegraphics[width=13.cm]{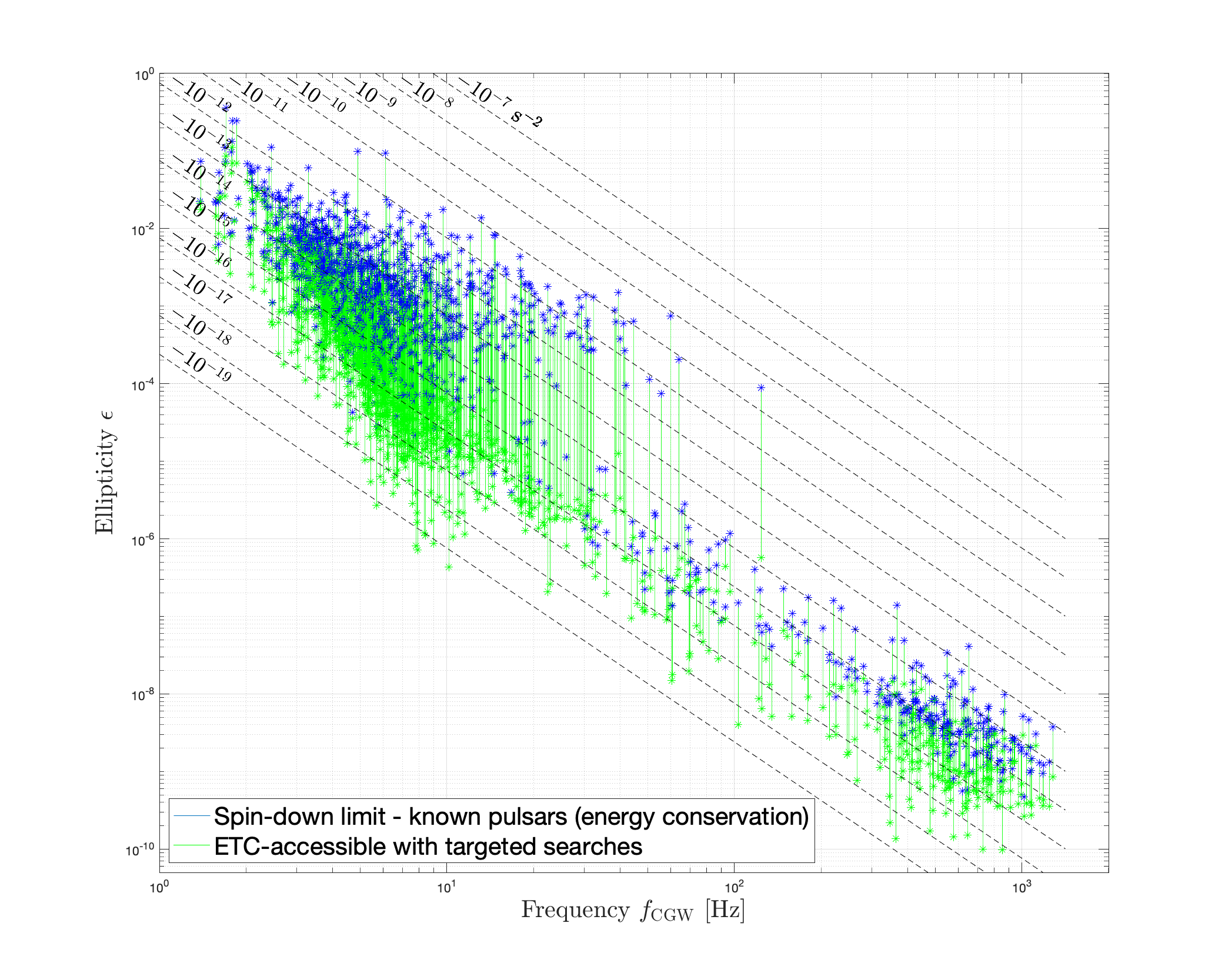}
\caption{Nominal expected GW frequencies and maximum allowed ellipticities for known
  pulsars.  Blue asterisks indicate ellipticities
  accessible with ETC sensitivity (3 detectors, 5 years), using targeted searches,
  where  green vertical lines and green asterisks indicate additional ellipticity sensitivity range
  for ETC.  Diagonal dashed lines show corresponding $\fcgwdot$ values under the gravitar model.}
\label{fig:fvsepsilon}
\end{center}
\end{figure}

Maximum possible upper limits from energy conservation
may have little to do, of course, with maximum physically possible ellipticities, must less to do
with physically {\it plausible} ellipticities. Figure~\ref{fig:fvsepsilon} shows maximum possible
ellipticities for known pulsars (blue asterisks) based on inferred $\fcgw$, $\fcgwdot$ and $d$ values
for which the ETC targeted-search strain sensitivity could probe those values. Vertical lines
terminated by green asterisks indicated the full range of ellipticity sensitivity achievable
with ETC. Sensitivities to very low
ellipticities come mainly from the highest-frequency stars, as expected from equation~\ref{eq:hexpected},
$h_0\propto \epsilon\fcgw^2$.
For example, no known pulsar with a maximum ellipticity below $10^{-7}$ and that is ETC-accessible has a $\fcgw$ value
lower than 90 Hz. and all ETC-accessible pulsars with a maximum ellipticity below $10^{-9}$ lie above 500 Hz.

\begin{figure}[t!]
\begin{center}
\includegraphics[width=13.cm]{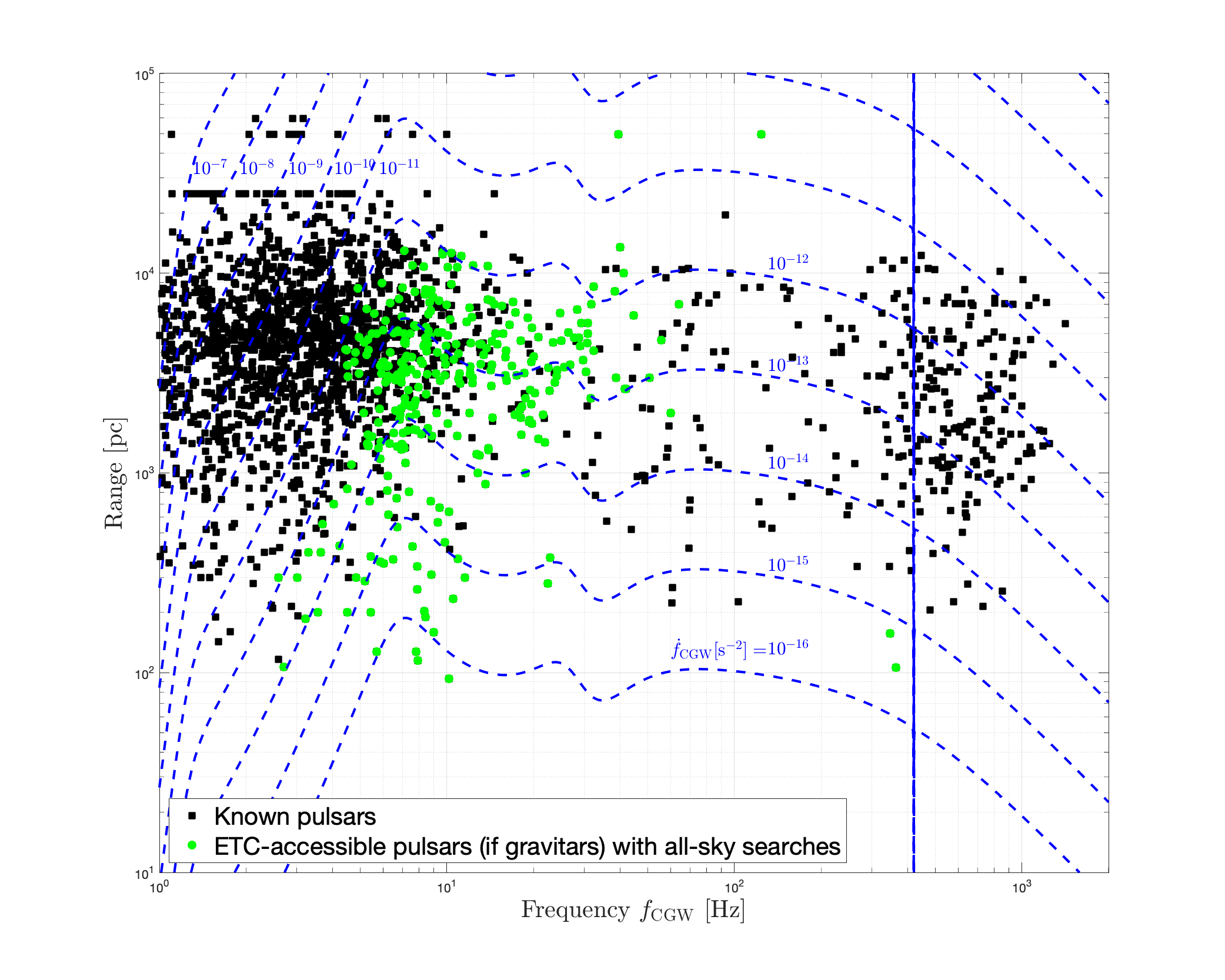}
\caption{Maximum allowed all-sky-search ranges for gravitars vs GW frequencies for different assumed
  spin-down derivatives for ETC sensitivity (dashed blue curves). Known pulsar
  distances are shown vs the expected GW frequencies, where green dots indicated pulsars with accessible spin-down limits for ETC sensitivity.
  These all-sky search ranges assume a sensitivity depth of 50 Hz$^{-1/2}$.}
\label{fig:fvsdist2}
\end{center}
\end{figure}

Since there is no guarantee that any of the known pulsars will be detectable with ETC sensitivity, one
can explore the possibility that a presently unknown neutron star, not presenting as a pulsar beaming toward
the Earth, is nonetheless close enough to us, spinning rapidly enough, and possessing a high enough
non-axisymmetry to be detectable, despite the reduced intrinsic strain sensitivity possible with
a computationally limited all-sky search. Figure~\ref{fig:fvsdist2} shows ranges (pc) 
\vs\ signal frequency for ETC all-sky sensitivity. Also shown are contours for different
assumed $\fcgwdot$ values under the gravitar assumption. These contours
are relevant because all-sky searches are constrained by their maximum spin-down range,
which governs computational cost. The dramatic improvement in noise expected for XG detectors with respect to
current advanced detectors lowers the $\fcgwdot$ range needed to reach the galatic center region~\citep{bib:RilesLRR}.
Known pulsars for which an all-sky search technique can reach the spin-down limit
are shown in green for reference.

%%%%%%%%%%%%%%%%%%%%%%%%%%%%%%%%%%%%%%%%%%%%%%%%%%%%%%%%%%%%
%%%%%%%%%%%%%%%%%%%%%%%%%%%%%%%%%%%%%%%%%%%%%%%%%%%%%%%%%%%%
%%%%%%%%%%%%%%%%%%%%%%%%%%%%%%%%%%%%%%%%%%%%%%%%%%%%%%%%%%%%
\section{The Inverse Problem \label{sect:inverse_problem}}

% R-modes: degeneracy; M/R; shape of phase diagram
% Mountains, non-detection: upper limits on B from magnetic mountains; from thermal mountains; degeneracy with breaking strain
% Mountains: detecection: degeneracy between magnetic and elastic; population; spin-down and brakng indices; M/R for r-modes

A detection of a CGW will be a significant milestone for GW astronomy.  Given the detection of many compact binary coalescence signals, the main significance of a CGW detection is likely to lie in what it tells us about the source itself, in particular the high density EoS, rather than about gravity.  This leads us to the so-called \emph{inverse problem}, \ie, what do we learn in the event of a successful CGW detection?

There has been relatively little work on this aspect of CGW physics, in large part because it is rather a difficult one; see \citet{lu_etal_23} for a discussion of some relevant issues.  What we can learn depends upon whether we are seeing emission from an r-mode or from a mountain.  We will consider each case in turn.  But before doing so, there is an even more basic question that needs to be addressed: how can we tell whether a detected CGW is from an r-mode versus a mountain?

If we have a measurement of the star's spin frequency $f_{\rm spin}$  from electromagnetic observations (\eg, it is seen as a radio pulsar), then the ratio of the CGW frequency $f_{\rm CWG}$ to $f_{\rm spin}$ will provide the necessary discriminant: we expect $f_{\rm CGW} / f_{\rm spin} = 2$ for mountains, while $f_{\rm CGW} / f_{\rm spin} \approx 4/3$ for r-modes.  (The departure of this ratio from $4/3$ is in fact important for the inverse problem, as will be described in Section \ref{sect:inverse_r-modes}.)    

If we do not have a measurement  of the spin frequency, there is, in general,  no obvious way of distinguishing CGW emission from an r-mode \vs\ emission from a mountain.  There is one exception, but requires rather special circumstances to apply.  If the spin-down of the star is driven entirely by CGW emission, and if the amplitude of the r-mode is constant in time, and if the ellipticity of the mountain is constant in time, then conservation of angular momentum leads to different and potentially measurable evolutions of the spin frequency with time.  This is most conveniently characterised by the star's \emph{braking index}, $n$, which requires measurement of the first and second time derivatives of $f_{\rm CGW}$:
\be
n \equiv \frac{f_{\rm CGW} \ddot f_{\rm CGW}}{\dot f_{\rm CGW}^2} .
\ee
For pure r-mode spin-down $n=7$, while for pure mountain spin-down $n=5$.  In practice, measurements of $\ddot f_{\rm CGW}$ may not be easy, as it requires a sufficiently large spin-down to occur over the duration of the CGW observation; see \citet{sj_22, sjm_23} for estimates of when this is possible.

In reality, there may well be significant magnetic dipole emission accompanying the CGWs.  This complicates the inverse problem, but progress can still be made, is one makes assumptions about how the magnetic torque scales with spin frequency.  These issues, and their impact on the inverse problem, are explored in \citet{lu_etal_23} and \citet{hua_etal_23}.

In fact, analysis of the frequency content of CGW signals is a rather subtle issue, as non-steady rotation (i.e. free precession; see e.g. \citet{ja_02}) complicates the story further.  We refer the reader to  \citet{jones_22} for a more detailed discussion.

%%%%%%%%%%%%%%%%%%%%%%%%%%%%%%%%%%%%%%%%%%%%%%%%%%%%%%%%%%%%
%%%%%%%%%%%%%%%%%%%%%%%%%%%%%%%%%%%%%%%%%%%%%%%%%%%%%%%%%%%%
\subsection{The Inverse Problem for r-modes \label{sect:inverse_r-modes}}

%There have in fact been a few claims of possible r-mode observations by electromagnetic means.  \citet{ll_19} have argued that plateaus seen in the X-ray light curves of the short gamma ray burst GRB090510 can be explained by r-mode emission following the formation of a rapidly spinning magnetised neutron star. 

In the case  that one can verify that the observed CGW is from r-modes (using the techniques mentioned above), there are two main things that can be done.

Firstly, the relation $f_{\rm CWG} / f_{\rm spin} = 4/3$ applies only in the idealised case of r-mode emission from a slowly spinning perfect fluid star modelled using Newtonian gravity.  In reality, all the sorts of physical refinements mentioned in Section \ref{sect:r-modes} will change this ratio slightly, as described in \citet{ioj_15}, who argued that the corrections from general relativity will likely be dominant, and can be accurately parameterised in terms of the stellar compactness $M/R$, something which, for a given $M$, depends upon the EoS.  The dependence of r-mode frequency on the stellar composition is therefore of great interest, and has attracted further attention in \citet{cj_18} and in \citet{gpc_23}.

In fact, we can already see how such arguments would proceed, as   \citet{sm_14a, sm_14b} have presented X-ray evidence for r-modes in two LMXB systems, both of known spin frequency.  The observed ratios $f_{\rm CWG}$ to $f_{\rm spin}$ were used first by  \citet{ajh_14} and then by  \citet{cj_18}, to constrain $M/R$.  A future CGW detection of r-mode emission from a star of known frequency would allow for further and more convincing measurements of this sort.  To be of maximal use in constraining the equation of state, a second independent combination of $M$ and $R$ would need to be measured, but this is more challenging.  

Note that the dependence of $f_{\rm CGW}$ on stellar parameters is good from the point of view of solving the inverse problem, but complicates the searches for such emission from NSs of known spin frequency; see \citet{cetal_19, lvk_21_O3_J0537}.  A recent study has examined this inverse problem in detail: building on the work of \citet{gpc_23}, including the existence of so-called ``universal relations'' between various stellar parameters, \citet{ghosh_23} has shown how one can obtain information on stellar compactness, moment of inertia, the EoS, and distance.

The second aspect of the inverse problem for r-modes concerns the instability window itself, \ie, the diving curve in the spin frequency--temperature plane between stability and instability.  Two representative curves were shown above, in Figures \ref{fig:jetal_22_fig_9} and \ref{fig:ketal_20_fig_4}.  As noted in Section \ref{sect:r-modes}, the location of this curve depends upon the sum of all dissipative processes that tend to damp the r-mode.  By locating points of definite stability, one can start to map out the curve.  There is a strong degeneracy here: the location of the curve at any given temperature depends upon \emph{the sum} of all dissipative processes.  This degeneracy is broken slightly if one can map out the curve over a wide range of temperature.  For instance,  a highly non-trivial shape of the form of Figure \ref{fig:ketal_20_fig_4} would point to resonances with superfluid modes as the dominant dissipation mechanism, with all other mechanisms being sub-dominant.

The fact that we \emph{don't} see r-mode emission from rapidly spinning stars, \ie, from millisecond pulsars and the NSs in LMXBs, already gives a constraint on the dissipation mechanism, but is a somewhat weaker one, as it simply tells us that the sum of all dissipation mechanisms is sufficient to stabilise the r-mode, without giving us individual constraints on each.

%%%%%%%%%%%%%%%%%%%%%%%%%%%%%%%%%%%%%%%%%%%%%%%%%%%%%%%%%%%%
%%%%%%%%%%%%%%%%%%%%%%%%%%%%%%%%%%%%%%%%%%%%%%%%%%%%%%%%%%%%
\subsection{The Inverse Problem for mountains \label{sect:inverse_mountains}}

The inverse problem for mountains is, unfortunately, much less easy than that for r-modes.  Assuming that a CGW signal is conformed as coming from a mountain (probably through measurement of the ratio $f_{\rm CGW} / f_{\rm spin} = 2$), that in itself provides no diagnostic between an elastic mountain versus a magnetic mountain.  One needs to appeal to additional ideas to try and make this distinction.

Before exploring such discriminants, let's examine the case of what one learns through the \emph{non-observations} of CGWs.  For a star of known spin frequency and known distance, the upper limit on GW amplitude translates into an upper limit on the ellipticity.  If one then assumes the actual (not detected) CGW  is purely from an elastic mountain, one can go a little further.  The ellipticity of equation (\ref{eq:epsilon_max_elastic}) is an estimate on the \emph{maximum} elastic mountain size, maximal as the crust is strained to its elastic limit.  The non-detection of emission at this level may simply reflect that the star does not happen to be strained at this level; it does not constrain the actual values of the rigidity parameter $b$ (the ratio of elastic binding energy of crust to gravitational binding energy), or the maximum breaking strain $u_{\rm max}$.  The value of the rigidity parameter is relatively well constrained by nuclear theory, to $b  \sim 10^{-5}$, so in practice the non-detection provides an upper limit on the actual strain $u$.  However, we do not have clear expectations of what sort of level of strain to expect, making it difficult to place this value in physical context.

If instead one assumes that the (not detected) CGW emission is purely from a magnetic mountain, the argument is simpler.  Via equations (\ref{eq:epsilon_B_normal}) or  (\ref{eq:epsilon_B_supercon}) above (depending upon whether the star is young and therefore hot enough to have a normal or superconducting core), one immediately gets an upper limit on the internal magnetic field strength.  Indeed, such arguments have already been used in interpreting non-detections of CGWs from known pulsars; see \citet{lvk_20_MSPs}, where the non-detection of CGWs from the Crab pulsar was used to bound the internal field as $B \lesssim 9 \times 10^{14}$ G.  To put this in context, the \emph{external} magnetic field strength of the Crab pulsar is inferred, via its observed spin-down rate, to be approximately $4 \times 10^{12}$ G, so that the CGW bound on the internal field strength is about $200$ times larger than its observed external field strength.  If the ET were to improve this upper bound by one order of magnitude, one could show the internal field is no more than a factor of $10$ larger than the external one, a significantly more interesting statement.

Turning back to the inverse problem for mountains following a CGW detection, the key difficulty is the one flagged at the start of this section: there is  no way, from the CGW signal alone, to distinguish between an elastic and a magnetic mountain.  One possibility is to convert the observed ellipticity into both a crustal strain $u$, or an internal magnetic field $B$, and see which seems more plausible.  In particular, the inferred internal field could be compared with (if it is known) the inferred external field strength; a very large discrepancy between the two would disfavour the magnetic mountain hypothesis relative to the elastic hypothesis.  However, the structure of magnetic fields within neutron stars is still not well understood (see, \eg, \citet{lj_12}), and until it is, it is not possible to make a quantitative  statement about how large a discrepancy between internal and external field strengths is required to rule out the magnetic mountain  explanation.
 
In the case of accreting stars, one can also consider the inverse problem for the case of the thermo-elastic mountains of \citet{oj_20} and \citet{hj_23}, to bound the internal magnetic field strength, but in this case one will obtain somewhat weaker bounds (\ie, larger maximum possible values) than for the conventional magnetic mountain case.

%%%%%%%%%%%%%%%%%%%%%%%%%%%%%%%%%%%%%%%%%%%%%%%%%%%%%%%%%%%%
%%%%%%%%%%%%%%%%%%%%%%%%%%%%%%%%%%%%%%%%%%%%%%%%%%%%%%%%%%%%
%%%%%%%%%%%%%%%%%%%%%%%%%%%%%%%%%%%%%%%%%%%%%%%%%%%%%%%%%%%%
\section{Conclusions \label{sect:conclusion}}

As we hope we have made clear, there is a large potential scientific return on the observation of a CGW, but there are several complicating issues.  Firstly, as described in Section 2, there is a large uncertainty in the likely strengths of CGW signals.  For CGWs from known pulsars, this is in large part a  reflection of our ignorance of the mechanism that would produce the required asymmetries in the NS's mass distribution, and the level of dissipation suffered by the r-mode instability.  In the case of CGWs from other classes of NS, including NSs in supernova remnants and those sought in all-sky searches, there is the additional uncertainty of the frequency of the CGW emission.  This means that one cannot make a confident statement about the likelihood of the detection of a CGW with an XG instrument.

We can at least quantify how our ability to make detections with an XG instrument depends upon NS parameters, as we have tried to do in Section \ref{sect:methodology}.  As made clear in our Figure \ref{fig:fvshspindown}, if one considers the energy-based upper bound on CGW emission from mountains on known pulsars, a substantial fraction of the known population are potentially detectable by the ET.  Looking in addition at our Figure \ref{fig:fvsepsilon},  we see that in the case of the young pulsars, the required ellipticities are uncomfortably high, given the discussion of crust shear modulus and breaking strain of Section \ref{sect:sources}, but the ellipticies required for the millisecond pulsars population are relatively modest, at the $\epsilon \sim 10^{-8}$ level or smaller, and potentially sustained by  ``normal" NS crusts.

Does this mean that the millisecond pulsars are a significantly better bet than young pulsars for detection by an XG instrument?  Not necessarily.  Young pulsars have a significantly different evolutionary history from the MSPs, so that different asymmetry-inducing mechanisms may apply to each.  Young neutron stars might gain asymmetries at birth, or as they spin-down rapidly soon after, while MSPs might gain asymmetries during their accretion phase.  Furthermore, if NSs do in fact turn out to have ``exotic cores" such that the strong nuclear force comes into play, the ellipticies of many younger more slowly spinning pulsars are potentially large enough to be interesting for CGW detection.  The ``eyes wide open'' strategy pursued in past and current CGW searches, of spreading efforts over a wide class of possibilities, still seems a sensible approach.

Given the uncertainties and difficulties outlined above, we suggest a few areas where more research would seem particularly useful: 
\begin{enumerate}
\item Further development of quantitative strategies to maximise the chances of CGW detections given our (rather uncertain) astrophysical knowledge and the (extremely severe) computational constrains of many CGW searches (see, for example, \citet{bib:MingEtalOptimization} and \citet{bib:ReedEtal}).
\item More modelling of \emph{likely} (rather than \emph{maximum}) CGW signal strengths.
\item More modelling of the likely instability parameter space for the r-mode.
\item More thought on the \emph{inverse problem} of how to extract information on NS physics and the properties of dense matter following a CGW detection.
\end{enumerate}
We hope that progress on these issues will help make CGW detection a reality, and an important tool in probing the properties of dense matter in the XG era.

%%%%%%%%%%%%%%%%%%%%%%%%%%%%%%%%%%%%%%%%%%%%%%%%%%%%%%%%%%%%
%%%%%%%%%%%%%%%%%%%%%%%%%%%%%%%%%%%%%%%%%%%%%%%%%%%%%%%%%%%%
%%%%%%%%%%%%%%%%%%%%%%%%%%%%%%%%%%%%%%%%%%%%%%%%%%%%%%%%%%%%
\section*{Acknowledgements}

The authors would like to thank their fellow members of the LIGO-Virgo-KAGRA Continuous Waves group for useful discussions, and feedback on this article.  DIJ acknowledges support from the Science and Technologies Funding Council (STFC) via grant No. ST/R00045X/1.  KR acknowledges support from National Science Foundation Award PHY-2110181.

\bibliography{merged_references}

\end{document}